\documentclass[aps,superscriptaddress,preprint]{revtex4}
\usepackage{graphicx}
\begin{document}
\title{Kinetic stabilization of Fe film on $(4\times 2)$-GaAs(100)}

\author{Jae-Min Lee}
\affiliation{School of Physics \& Center for Strongly Correlated
Material Research, Seoul National University, Seoul 151-747,
Korea}
\author{S.-J. Oh}
\affiliation{School of Physics \& Center for Strongly Correlated
Material Research, Seoul National University, Seoul 151-747,
Korea}
\author{K.J. Kim}
\affiliation{Pohang Accelerator Laboratory, Pohang University of
Science and Technology, Pohang 790-784, Korea}
\author{S.-U. Yang}
\affiliation{Department of Physics, Sookmyung Women's University,
Seoul 140-742, Korea}
\author{J.-H. Kim}
\affiliation{Department of Physics, Sookmyung Women's University,
Seoul 140-742, Korea}
\author{J.-S. Kim}
\affiliation{Department of Physics, Sookmyung Women's University,
Seoul 140-742, Korea}

\date{\today}

\begin{abstract}

We grow Fe film on $(4\times 2)$-GaAs(100) at low temperature,
($\sim130$ K) and study their chemical structure by photoelectron
spectroscopy using synchrotron radiation. We observe the effective
suppression of As segregation and remarkable reduction of alloy
formation near the interface between Fe and substrate. Hence, this
should be a way to grow virtually pristine Fe film on GaAs(100).
Further, the Fe film is found stable against As segregation even
after warmed up to room temperature. There only forms very thin,
$\sim 8 \AA$ thick interface alloy. It is speculated that the
interface alloy forms via surface diffusion mediated by interface
defects formed during the low temperature growth of the Fe film.
Further out-diffusion of both Ga and As are suppressed because it
should then proceed via inefficient bulk diffusion.

\end{abstract}

\maketitle

Fe film on GaAs(100) has been extensively studied as a
representative system for ferromagnetic metal-semiconductor
heterostructure.\cite{Prinz} Due to the small lattice mismatch
($\sim 1.3\%$) between the double of the lattice constant of Fe
and that of GaAs, epitaxial growth of Fe film on GaAs(100) is
achieved. However, alloy formation near the interface and serious
outdiffusion of both Ga and As from the bulk have been notorious
problems.\cite{Ruckman,Chambers,Kneedler} There have been various
attempts to solve those problems such as S-passivation of GaAs
surface \cite{Anderson} and insertion of Er layer between Fe and
GaAs\cite{Schultz}. For both cases, the interfacial reaction is
reduced to some extent,  but the segregated As is still observed.
Chye {\it et al.}\cite{Chye} grow Fe film on GaAs (100) at 120 K
and also insert Al interlayer. They find improved squareness in
magnetic hysteresis and reduced interface states in
photo-luminescence spectra. However, no direct investigation on
atomic and chemical structure of the Fe film is made.

In the present work, we grow Fe film on GaAs(100) around 130 K,
and examine the possibility of kinetic stabilization of the Fe
film by photoelectron spectroscopy. Here, we report direct
evidence for the effective suppression of the outdiffusion of both
Ga and As and the minimal formation of the interface alloy during
the growth of the Fe film. When the film is warmed up to room
temperature, there forms very thin alloy limited near to the
interface. Further outdiffusion of both Ga and As is, however,
still suppressed.
\par

All the experiments are performed at 2B1 beam line of Pohang light
source in Korea. It is equipped with an electron energy analyzer
and low energy electron diffraction (LEED) optics. The base
pressure of the chamber is below $5\times 10^{-10}$ Torr. For the
substrate, we use Si-doped, n-type GaAs(100). Repeated sputtering
and annealing produce clean and well-ordered GaAs substrate;
sputtering is made by Ar ion beam of 0.5 K eV, with its incidence
angle 45$^o$ from surface normal to minimize surface damage.
Annealing is made at 840 K for 30 minutes. As-prepared surface
shows well-defined 4 $\times$ 2 LEED pattern.
\par

We use an e-beam evaporator to deposit Fe film whose thickness is
determined by a quartz microbalance that is calibrated by {\it in
situ} surface x-ray reflectivity (SXR) measurement for thick Fe
films. The deposition rate is 0.45 $ \AA $  per minute. Temperature
of the sample is determined by both an N-type thermocouple and a
Si-diode attached near to the sample. Both the deposition of the Fe
film and the acquisition of photoelectron spectra are made at the
same temperature. All the spectra are taken with a beam energy, 100
eV. Most of the spectra are taken in normal emission geometry,
unless otherwise stated.
\par

In Fig. 1 and 2, shown are the variations of the spectra for Ga 3d
and As 3d with increasing Fe thickness. Each spectrum is
normalized to have the same maximum intensity, and vertically
shifted for the clear presentation. Binding energies of Ga 3d and
As 3d spectra are relative to that of the respective 3d$_{5/2}$
peak. For the Fe film grown on the substrate at room temperature
(RT-grown sample), the peaks for Ga and, especially As 3d are
still evident even for the Fe films respectively of 20 $\AA$ for
Ga and 40 $\AA$ for As that are much thicker than the escape
lengths of the relevant photoelectrons, $\sim$ 3 $\AA$. It
indicates that substantial amount of Ga and As is segregated from
the bulk  as previously reported.\cite{Ruckman,Chambers,Kneedler}
In sharp contrast to the RT-grown sample, for the Fe film grown at
low substrate temperature, around 130 K, (LT-grown sample), very
rapid decrease of the peak intensities for both Ga and As 3d is
observed as the coverage of the Fe film increases.
 Such an observation strongly suggests that for the LT-grown sample
the out-diffusion of Ga and As should be effectively suppressed.\par

For the quantitative information on the chemical structure of the
system, we further analyze the spectra: Each spectrum can be fit
by three components, each of which is formed of spin-orbit split
two peaks ($3d_{5/2}, 3d_{3/2}$). (Fig. 3) Similar scheme of
fitting was also made by Ruckman {\it et al.}\cite{Ruckman} and
Kneedler {\it et al.}\cite{Kneedler}. One difference from their
analyses is that Ga 3d spectrum is now fit by three components
instead of two, because a broad feature observed in the previous
works with its relative binding energy around -0.5 $\sim$ 1 eV is
now well resolved into two peaks due to improved resolution in the
present work (Fig. 1).
In the fitting process, we held the Gaussian widths around
0.3$\sim$0.35 eV, but vary Lorentzian width.
As an example, we show the fitting results for Ga 3d and As 3d spectra
for an Fe film with its thickness of
3$\AA$, in the inset of Fig. 3 and 4, respectively. The energy difference
between $3d_{5/2}$ and $3d_{3/2}$ is about 0.45 eV for Ga and 0.69
eV for As. The relative binding energies for the three components of As ,
0, $\sim$ 0.3 and $\sim$
0.6 eV, are similar to the
corresponding values, 0, $\sim$ 0.4 and 0.7 eV, determined by Ruckman
{\it et al.}\cite{Ruckman}. Those of Ga are slightly different
from those of Ruckman {\it et al.} due to the increased number of
peaks employed for the present fitting as mentioned above.
\par

In Fig.3, summarized is the peak area for each component of Ga 3d
as a function of the thickness of the Fe film. For both the RT-
and LT-grown samples, the peak areas of the two components(denoted
as bulk and reacted 1.) of Ga 3d decreases exponentially. The
remaining one (denoted as reacted 2.) initially increases for Fe
thickness up to $\sim 8 \AA$, and then decreases for the RT-grown
sample. From those observations, we attribute the first two peaks
to the bulk Ga and the Ga adjacent to the Fe film, while the
remaining one to Ga forming solid solution with Fe near the
interface or interface alloy. A notable difference between the
RT-growth and the LT-growth is observed for the interface alloy.
In contrast to the RT-grown sample, that component never shows
sizable intensity for the LT-grown sample. (Fig. 3) In short, for
the RT- grown sample, Ga forms interface alloy with Fe of
thickness, $\sim 8\AA$ while it is almost absent for the LT-grown
sample.
\par

 Fig.4 shows the peak areas for the three components of As
as a function of the thickness of Fe film, respectively for the
RT- and LT- grown samples. As for the case of the Ga 3d, for the
RT-grown sample, the peak areas of the two components (denoted as
bulk and reacted 1.) show exponential decay as the thickness of
the Fe film increases. In the thick film limit, the bulk component
decreases to show nearly null intensity. That for the reacted 1
reaches a non-zero, but quite low intensity, and tentatively
assigned to a solid solution phase of As with Fe film. On the
other the remaining one (denoted as the reacted 2.) persists with
very large intensity ($\sim$ 26$\%$ of that of bare substrate)
even for the Fe film thicker than 40 $\AA$, and is considered as
surface segregated As. This component shows enhanced intensity
when the spectrum is taken in 60$^o$ off-normal geometry (figure
not shown), consistently with the assignment of the component to a
surface segregated As. Such segregated As has been previously
reported by other groups.\cite{Ruckman,Chambers,Kneedler}

For the LT-grown sample, however, the component corresponding to
the segregated As shows disparate behavior or exponential decay of
its peak area after initial, short-term (up to 3$\AA$ of the Fe
film) increase. (Fig. 4) The decay length of the total peak area
is, however, only $\sim$$2 \AA$, indicating that the outdiffused
As is less than 1 ML in regards to the escape depth of the As 3d,
$\sim 3\AA$. Such an experimental observation tells that the
out-diffusion of As can be virtually suppressed by the growth of
the Fe film at low substrate temperature, which is here $\sim$ 130
K.\par

Although the surface segregation and interfacial alloying of Ga
and As is dramatically reduced by the LT-growth of Fe film, the
thermal stability of the LT-grown sample is an important issue,
considering its application for devices whose working temperatures
should be mostly around room temperature. For the examination of
the thermal stability of the LT-grown Fe film, a LT-grown sample,
30$\AA$ thick Fe film, is gradually warmed up to room temperature.
As 3d peak shows little intensity discernable from the background
noise. (Inset of Fig. 2) Moreover, its intensity does not show any
enhancement in 60$^o$ off-normal emission geometry (figure not
shown) either. Thus, we conclude that As does not segregate to the
surface at room temperature. Since the probing depth of the
photoelectron is limited near to the surface, we still do not know
what happens below the surface region. Preliminary results of our
surface x-ray reflectivity study\cite{sxr} finds that upon warming
a LT-grown sample, 23$\AA$ Fe on GaAs(100), to room temperature,
there newly forms $\sim 8\AA$ thick interface alloy. It is
speculated that during the annealing process, some substrate atoms
diffuse through the vacancies or pinholes near the interface which
form in the Fe layers during the low temperature growth. After
defects near the interface are exhausted by the substrate atoms,
further out-diffusion of them is inhibited, because it should then
proceed via bulk diffusion which is much less efficient than the
surface diffusion, and hence the LT-grown Fe film is kinetically
stabilized.\par

We grow Fe film on GaAs(001) around 130 K, to kinetically suppress
the out-diffusion and the interface alloying of both Ga and As.
High resolution photoelectron spectroscopy finds that the
interface alloying is limited within several mono-layers from the
interface, and the out-diffusion of Ga and especially As is almost
suppressed. Such virtually pristine Fe film seems preserved even
upon warming the sample up to room temperature, since little
spectroscopic change is observed. The thermal stability of the Fe
film has kinetic origin; the outdiffusion of the substrate atoms
to the Fe film should proceed via bulk diffusion. The kinetically
stabilized pure Fe film on GaAs(100) can be widely applicable for
the fabrication of devices using ferromagnetic metal-semiconductor
junction, especially for the efficient spin
injectors.\cite{Ploog,Hanbicki2003}\par

{\it Acknowledgement}-This work is supported by KOSEF through CSCMR and by Pohang Accelator
laboratory.

\newpage

\begin{figure}
\includegraphics[width=0.9\textwidth ]{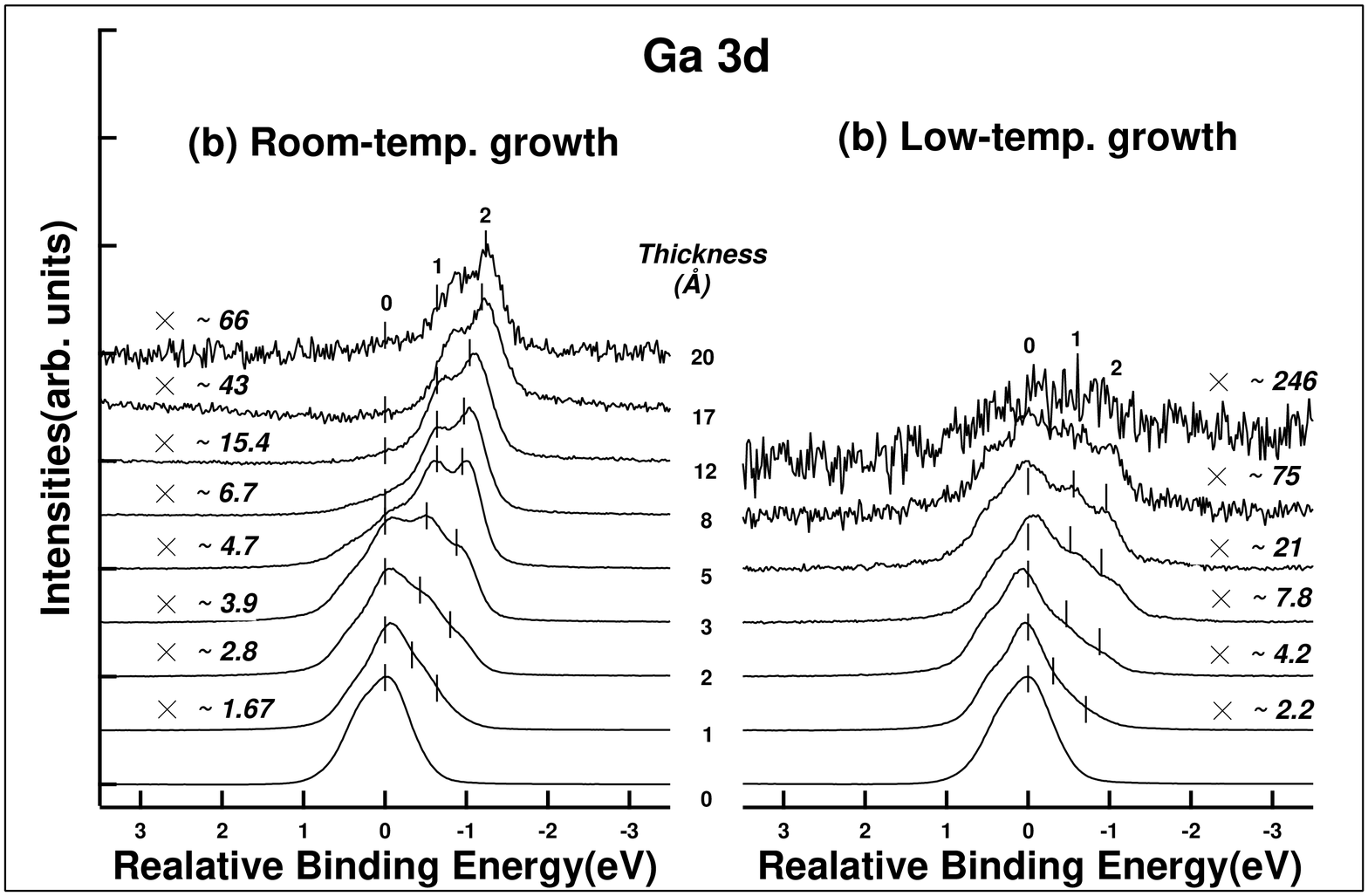}
\caption[] {Ga 3d spectra in the order of increasing Fe film
thickness. Growth temperature of the Fe film, (a): room temperature,
(b): $\sim$130 K. } \label{fig:1}
\end{figure}

\begin{figure}
\includegraphics[width=0.9\textwidth ]{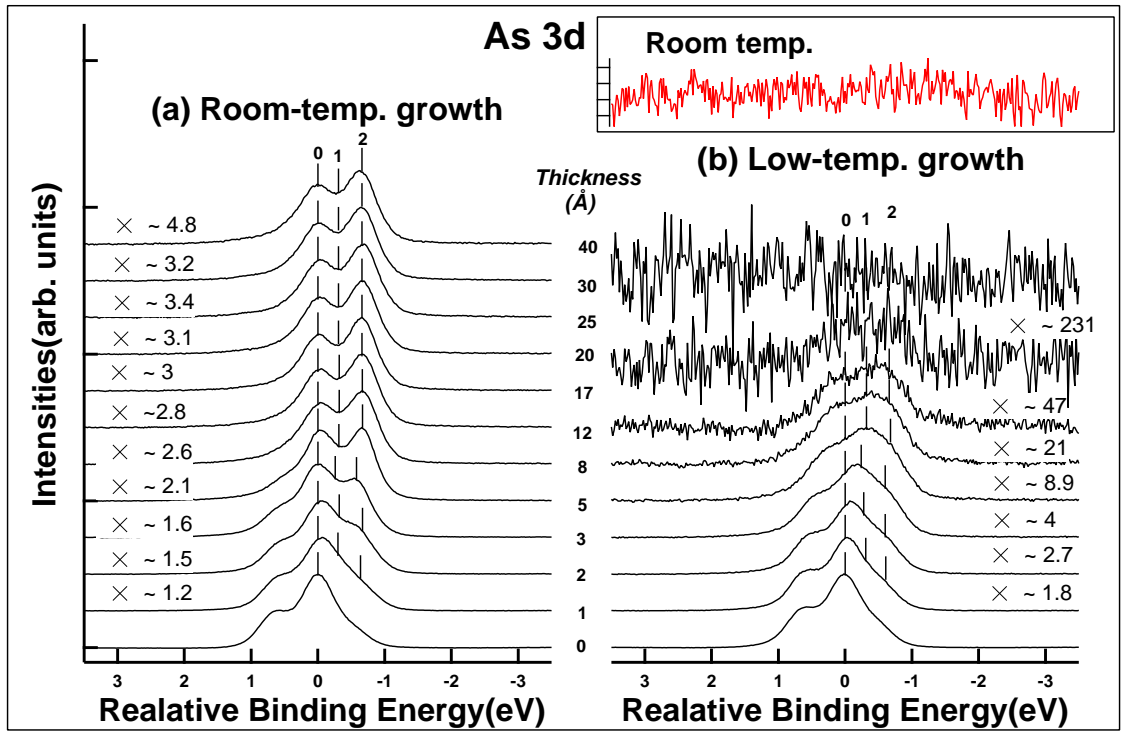}
\caption[] {As 3d spectra in the order of increasing Fe film
thickness. Growth temperature of the Fe film, (a): room temperature,
(b): $\sim$ 130 K. Inset: An As 3d spectrum for a LT-grown sample
when it is annealed up to 300K.) } \label{fig:2}
\end{figure}


\begin{figure}
\includegraphics[width=0.9\textwidth]{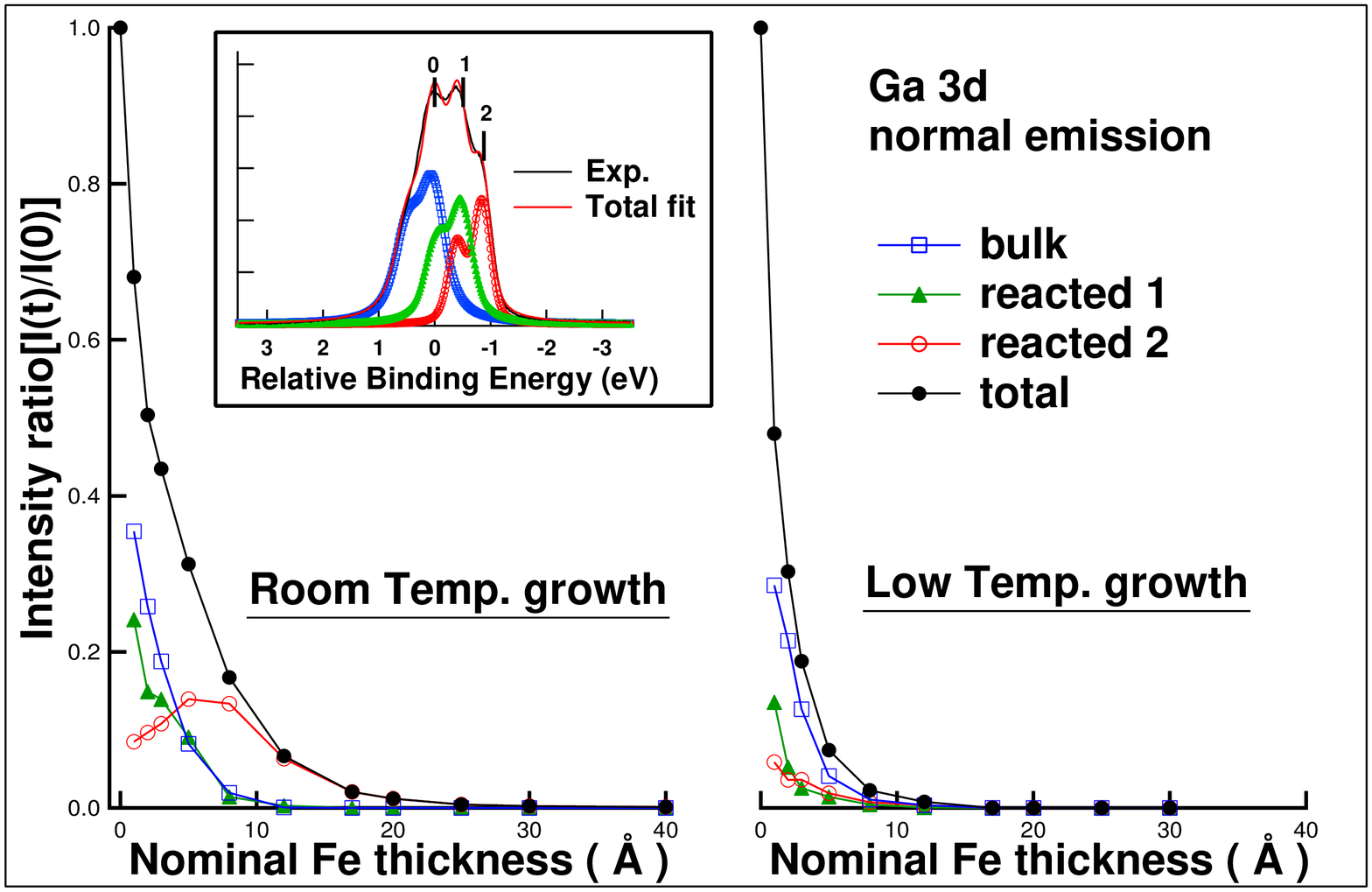}
\caption[] {(Color online) Peak areas of the three componenets of Ga
3d vs the thickness of the Fe film grown (a) at room temperature and
(b) low temperature, $\sim$ 130 K. Inset: Best-fit curves for a Ga
3d spectrum (background subtracted) for a sample with 3$\AA$ Fe
grown on GaAs(100) at room temperature.} \label{fig:4}
\end{figure}


\begin{figure}
\includegraphics[width=0.9\textwidth]{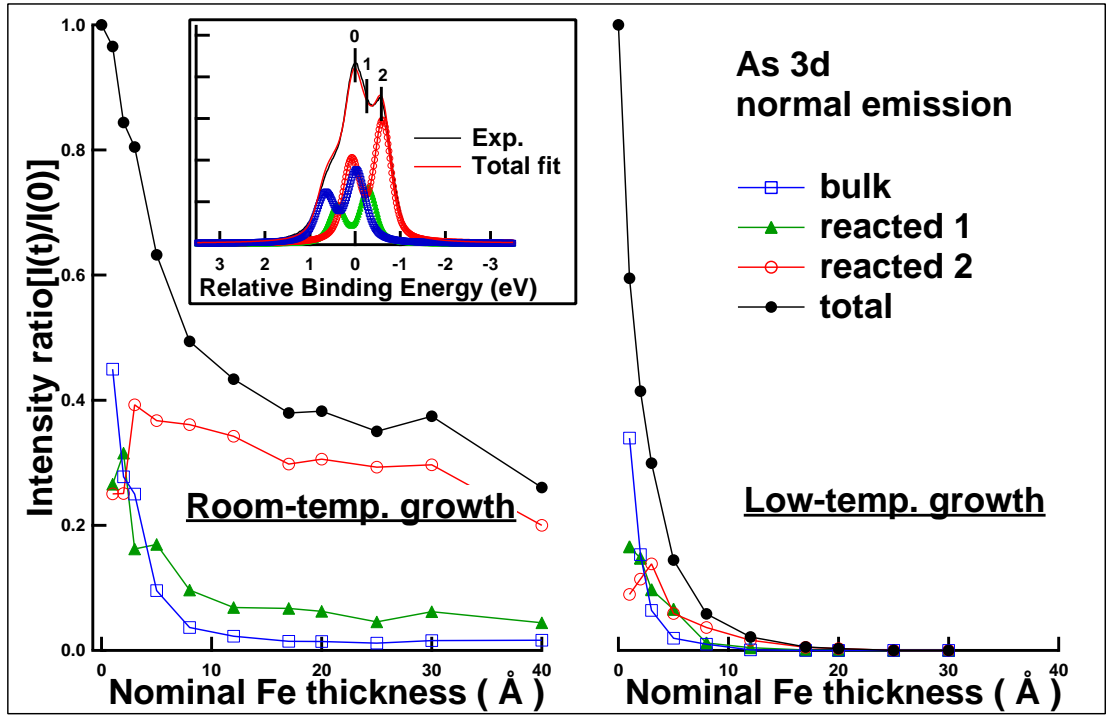}
\caption[] {(Color online) Peak areas of the three components of As
3d vs the thickness of the Fe film grown (a) at room temperature and
(b) at low temperature, $\sim$ 130 K. Inset: Best-fit curves of a As
3d spectrum(background subtracted) for a 3$\AA$ thick Fe film on
GaAs(100) grown at room temperature.} \label{fig:5}
\end{figure}


\end{document}